\documentclass[12pt]{iopart}
\usepackage{amsfonts,epsfig}
\def\kperp{k_\perp}
\def\avet{\overline{E}_{T}}

\def\bq{\begin{equation}}
\def\eq{\end{equation}}
\def\ba{\begin{eqnarray}}
\def\ea{\end{eqnarray}}

\def\mboxsc#1{\mbox{\scriptsize #1}}
\def\smin{s_{\mbox{\scriptsize min}}}
\def\ycut{y_{\mbox{\scriptsize cut}}}

\def\Journal#1#2#3#4{{#1} #2 (#4) #3}
\def\NPB{{Nucl.~Phys.} B}
\def\PLB{{Phys.~Lett.}  B}

\def\PRD{{Phys.~Rev.} D}
\def\ZPC{{Z.~Phys.} C}
\def\EPJC{{Eur.~Phys.~J.} C}
\newcommand{\jv}{{\tt JetViP}}
\def\cm{\checkmark}

\begin{document}
\thispagestyle{empty}

\begin{center}
\centerline{\hfill  MPI/PhT/2001--039}
\centerline{\hfill  September 2001}
\vskip1.5cm
\renewcommand{\thefootnote}{\fnsymbol{footnote}}
{\huge Jet Production at HERA\footnote[5]{Invited talk at the
Ringberg Workshop {\it New Trends in HERA Physics 2001}, Ringberg
Castle, Tegernsee, Germany, 17--22 June 2001}} 
\renewcommand{\thefootnote}{\arabic{footnote}}
\vskip1.cm
{\Large B.~P\"otter \\
Max-Planck-Institut f\"ur Physik (Werner-Heisenberg-Institut),\\
F\"ohringer Ring 6, 80805 Munich, Germany \vspace{2mm} \\
e-mail: poetter@mail.desy.de }

\vspace{2cm}

{\Large Abstract} 

\vspace{.7cm}

\begin{minipage}[t]{11cm} 
I review the status of perturbative QCD calculations for jet
production in $eP$-scattering at HERA. I will discuss
possibilities of combining fixed order, especially higher order,
calculations and parton showers.
\end{minipage}
\end{center}
\cleardoublepage
\setcounter{page}{1}

\title[Jet Production at HERA]{Jet Production at HERA}

\author{Bj\"orn P\"otter\footnote[3]{Present address: European
Aeronautic Defence and Space 
Company (EADS), Munich. Mail to poetter@mail.desy.de}}

\address{Max-Planck-Institut f\"ur Physik, F\"ohringer Ring 6, 80805
Munich, Germany}

\begin{abstract}
I review the status of perturbative QCD calculations for jet
production in $eP$-scattering at HERA. I will discuss
possibilities of combining fixed order, especially higher order,
calculations and parton showers.
\end{abstract}

%**********************************************************************
%**********************************************************************
\section{Introduction}

Jet production in high-energy scattering is a classical testing
ground for QCD. Not only can one measure typical QCD quantities, such
as the strong coupling $\alpha_s$ or the parton distributions
functions (PDF's), but also one has a handle to test perturbative QCD,
including the factorization theorems. Furthermore jet production
processes can provide important backgrounds for the search of new
physics. Therefore jet production has recieved much attention in theoretical
calculations and impressive progress has been made in the last decade
to describe jet production in the framework of perturbative QCD.

Most available calculations have been performed in next-to-leading
order (NLO) accuracy. There are several reasons to perform these NLO
calculations. First, the theoretical uncertainties due to unphysical
renormalization and factorization scale dependences are
reduced. Second, due to the emission of additional particles in the
initial and final state, one becomes sensitve to jet algorithms, which
is certainly the case in the experimental results. Third, for the
same reason, calculations become more sensitve to detector
limitations. Finally, the presence of infrared (IR) logarithms is clearly
seen and regions where resummation is needed can be identified. 

In the following I will review the state-of-the-art for
perturbative calculations for $eP$-scattering which can be tested at
HERA. Jet production in $eP$-scattering involves large
transverse energies $E_T$ or photon virtualities $Q^2$. The presence
of a large scale ensures that perturbative calculations can be
performed and one can hope that hadronization corrections and 
theoretical uncertainties are small.\footnote{In looking at
theoretical errors, one should keep in mind also the selection of a
stable jet algorithm to obtain reliable results \cite{potsey}.}
In the following I will not present many experimental results, since
these have been reviewed nicely in this workshop by M.~Wing \cite{wing}.

Presently the limiting factor for higher precision measurements of QCD
parameters are the systematic and theoretical uncertainties. Therefore, 
theoretical advances are in urge, such as NNLO calculations or NLO
Monte Carlo programs (MC's) including parton showers and hadronization
corrections. Therefore,  I will also discuss possibilities of
combining fixed order calculations and parton showers (PS), especially
the problem of incorporating higher order corrections in the fixed
order part of the MC's.

%**********************************************************************
%**********************************************************************
\section{Jet Cross Sections at Next-to-Leading Order}

I start by summarizing the procedure for the numerical evaluation of
an inclusive $n$-jet cross section in NLO QCD. The first step is to
select a jet algorithm, which defines how partons are recombined to
give jets. In the following we take for definiteness the invariant 
mass $s_{ij}$ of two partons $i$ and $j$ and define the $n$-jet
region such that $s_{ij}<\smin$, with some kind of minimum mass
$\smin$ and likewise the $(n+1)$-jet region such that $s_{ij}>\smin$ for
all $i,j$. The LO process for the production of $n$ jets consists of
$n$ final state partons and obviously does not depend on the jet
definition. This dependence only comes in at NLO. The ${\cal O}(\alpha_s)$ 
corrections to this process are given by the ultraviolet (UV) and
infrared (IR) divergent one-loop contributions to the $n$-parton
configuration, which are the virtual corrections, and the NLO tree
level matrix elements with $(n+1)$ partons, the real corrections. The
tree-level matrix elements have to be integrated over the phase space
of the additional parton, which gives rise to collinear and soft
singularities. After renormalization, the singularities in the virtual
and soft/collinear contributions cancel and remaining poles are
absorbed into parton distribution functions. One wants to integrate
most of the phase space of the real corrections numerically, but one
needs to find a procedure to calculate the soft/collinear
contributions analytically as to explicitly cancel the poles from the
virtual corrections. The two basic methods to perform these
integrations are the subtraction method  \cite{ERT,KS,CS} and the
phase-space slicing (PSS) method \cite{pps2,pps3,pps1,Gr} (see also
\cite{kramer} for a review).

In the following we will make use of the PSS method and therefore
discuss this method further. To illustrate the method, we rely on the
classical example given by Kunszt and Soper \cite{KS}. We label the LO
Born contribution as $\sigma^{\mboxsc{LO}}=\sigma^B$. The NLO cross
section is given by the sum of the Born cross section and the virtual
and real corrections, $\sigma^V$ and $\sigma^R$:
\bq
  \sigma^{\mboxsc{NLO}}=\sigma^B + \sigma^V + \sigma^R 
   = \sigma^B + C_V - \lim_{\epsilon\to 0} \frac{1}{\epsilon}F(0) +
   \int_0^1 \frac{dx}{x}F(x) \ . \label{KS-ex}
\eq 
Here, $F(x)$ is the known, but complicated function representing the
$(n+1)$-parton matrix elements. The variable $x$ represents an angle
between two partons or the energy of a gluon, the integral represents
the phase-space intergation that has to be performed over the
additional parton. The singularity of the real corrections at $x\to 0$ is
compensated by the virtual corrections, given by the pole term and some
constant, $C_V$. In the PSS method, the integral over the real
corrections is divided into two parts, $0<x <\delta$ and 
$\delta <x<1$. We note that the technical cut-off $\delta$ should
lie within the $n$ jet region, i.e., if we define
$\ycut=\smin/Q^2$, then we should have $\delta<\ycut$. If the
cut-off parameter is sufficiently small, $\delta \ll \ycut <1$, one
can write 
\begin{eqnarray}
  \sigma^R &=& \int_0^1 \frac{dx}{x}F(x) \simeq 
 \lim_{\epsilon \to 0} \left\{
 \int^1_{\delta} \frac{dx}{x} x^\epsilon F(x)
 +F(0) \int^{\delta}_0 \frac{dx}{x} x^\epsilon \right\} \nonumber \\ 
  &\simeq & \int^1_{\delta} \frac{dx}{x} F(x) +F(0) \ln(\delta) +
  \lim_{\epsilon\to 0} \frac{1}{\epsilon}F(0) \ ,
\end{eqnarray}
where the integral has been regularized by the term $x^\epsilon$, as
suggested by dimensional regularization. The pole is now explicit and
the NLO cross section $\sigma^{\mboxsc{NLO}}$ is finite:
\bq
  \sigma^{\mboxsc{NLO}} \simeq \sigma^B + C_V + \int_{\delta}^1
  \frac{dx}{x} F(x) + F(0)\ln(\delta) \ .  \label{ex1}
\eq
Clearly, the real corrections $\sigma^R$ should not depend on $\delta$,
and the logarithmic $\delta$ dependence of the last term in eqn
(\ref{ex1}) should be canceled by the integral, which sometimes is
numerically difficult for very small parameters $\delta$. However, an
improvement of the above solution is possible by using a hybrid of the
PSS and the subtraction methods, suggested by Glover and Sutton
\cite{GSu}. In this method, one adds and subtracts only the universal
soft/collinear approximations for $x < \delta$, such that 
\begin{eqnarray}
 \sigma^R &= & \lim_{\epsilon \to 0} \left \lbrace
 \int^1_0 \frac{dx}{x} x^\epsilon F(x)
  -F(0) \int^{\delta}_0 \frac{dx}{x} x^\epsilon
  +F(0) \int^{\delta}_0 \frac{dx}{x} x^\epsilon
  \right \rbrace \nonumber \\
    &\simeq & \int^1_{\delta} \frac{dx}{x} F(x) 
   + \int^{\delta}_0 \frac{dx}{x} \bigg[ F(x)-F(0) \bigg] 
   + F(0) \ln(\delta) +
     \lim_{\epsilon\to 0} \frac{1}{\epsilon}F(0)  \ .
\end{eqnarray}
A cancellation between the analytical and numerical terms still
occurs, however only the phase space is approximated, so that this
method is valid at larger values of $\delta$. In the case where the
phase-space is not approximated for small $x$ the hybrid method
becomes independent from $\delta$.

%**********************************************************************
%**********************************************************************
\section{Jet Production in $eP$-Scattering}

In $eP$-scattering, the interaction of the electron with the
proton is mediated by a gauge boson ($\gamma, Z^0, W^\pm$) with
virtuality $Q^2\ge 0$. The region from $Q^2=0$ (photoproduction) up to
the highest $Q^2>10^4$~GeV$^2$ (deep-inelastic scattering, DIS) is
covered by the HERA collider. In the DIS region there are four
available programs, namely {\tt DISENT} \cite{1}, {\tt DISASTER++}
\cite{2}, {\tt MEPJET} \cite{3} and \jv\ \cite{4,jv2}, based on the
calculations in \cite{jv-c}. The programs are summarized in Table~1 (taken
from \cite{comp}). So far, only MEPJET incorporates contributions from
$Z^0$ and $W^\pm$ exchange in NLO, which become important at
virtualities above $Q^2>2500$~GeV$^2$. Therefore, a detailed
comparison of the existing fixed order MC's is necessary to see
whether these contributions are reliably predicted (see \cite{jv2,comp}).

\begin{table}
\renewcommand{\arraystretch}{1.1}
\begin{center}
{\small
\begin{tabular}{@{}|l|llll|}
\hline
&MEPJET &DISENT &DISASTER++ &JETVIP \\
\hline 
version & 2.2 & 0.1 & 1.0.1 & 2.1\\
\hline \hline
method & PS slicing & subtraction & subtraction & PS slicing \\ \hline
%1+1     & NLO     & NLO    & NLO    & --- \\
1+1,2+1     & NLO     & NLO    & NLO    & NLO \\
3+1     & LO      & LO     & LO     & LO \\
4+1     & LO      & ---    & ---    & --- \\
\hline
full event record & \cm & \cm & \cm & \cm \\
\hline
flavour dependence & switch & switch & full & switch \\
\hline
Quark Masses          & LO  & --- & ---  & --- \\
Resolved $\gamma$       & --- & --- & ---  & NLO \\
Electroweak  & NLO & --- & ---  & --- \\
Polarized $e$/$P$ & NLO & --- & --- & --- \\
\hline
\end{tabular}}
\caption{Summary of $eP\to \mbox{jets}$ fixed order computer programs
in DIS.}
\end{center}
\end{table}

One of the main features of {\tt JetViP} is the possibility to include
a resolved virtual photon component in NLO. In this way the
photoproduction limit can be taken. There are several calculations
available for photoproduction \cite{Frixione:1997ks,Harris:1997hz,% 
Klasen:1996it}. The NLO corrections to the direct process in DIS become
singular in the limit $Q^2\rightarrow 0$ in the initial state on the
real photon side. For $Q^2=0$ these photon initial state singularities
are usually evaluated with the dimensional regularization
method. Then the singular contributions appear as poles in $\epsilon =
(4-d)/2$ with the form  $-\frac{1}{\epsilon}P_{q\gamma}$ multiplied
with the LO matrix elements for quark-parton scattering. These singular
contributions are absorbed into PDF's $f_{a/\gamma }(x)$ of the real
photon. For $Q^2\neq 0$ the corresponding contributions are replaced
by \cite{jv-c}.   
\begin{equation}
 -\frac{1}{\epsilon} P_{q\gamma} \rightarrow -\ln(s/Q^2) P_{q\gamma}
\end{equation}
where $\sqrt{s}$ is the c.m. energy of the photon-parton subprocess. These
terms are finite as long as $Q^2 \neq 0$ and can be evaluated with $d=4$
dimensions, but become large for small $Q^2$, which suggests
to absorb them as terms proportional to $\ln(M_{\gamma }^2/Q^2)$ in the
PDF of the virtual photon. Parametrizations of the virtual photon have
been provided by several groups \cite{vph}. By this absorption the PDF of
the virtual photon becomes dependent on $M_{\gamma}$, which is the
factorization scale of the virtual photon, in analogy to the real
photon case. Of course, this absorption of large terms is necessary
only for  $Q^2 \ll M_{\gamma}^2$. In all other cases the direct cross
section can be calculated without the subtraction and the additional
resolved contribution. $M_{\gamma}^2$ will be of the order of
$E_T^2$. But also when $Q^2 \simeq M_{\gamma}^2$, one can perform this
subtraction. Then the subtracted term will be added again in the
resolved contribution, so that the sum of the two cross sections
remains unchanged. In this way also the dependence of the cross section
on $M_{\gamma }^2$ must cancel, as long as the resolved contribution 
is calculated in LO only.

%**********************************************************************
%**********************************************************************
\section{Jet Cross Sections at Low $Q^2$}  

\begin{figure}[ttt]
\centering
 \epsfig{file=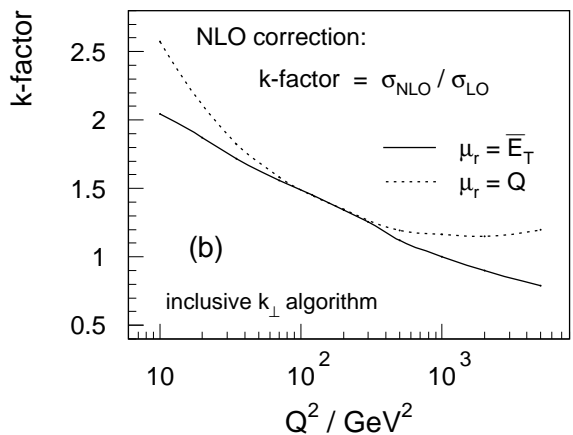,width=7.95cm}
\vskip-3mm
\caption{The predictions of the next-to-leading order corrections to
the dijet cross section as a function of $Q^2$ for the inclusive
$\kperp$ algorithm using two different renormalization scales $\mu_r$.}
\label{fig:hadcoralgo}
\end{figure}

\begin{figure}[bbb]
\begin{center}
~\epsfig{file=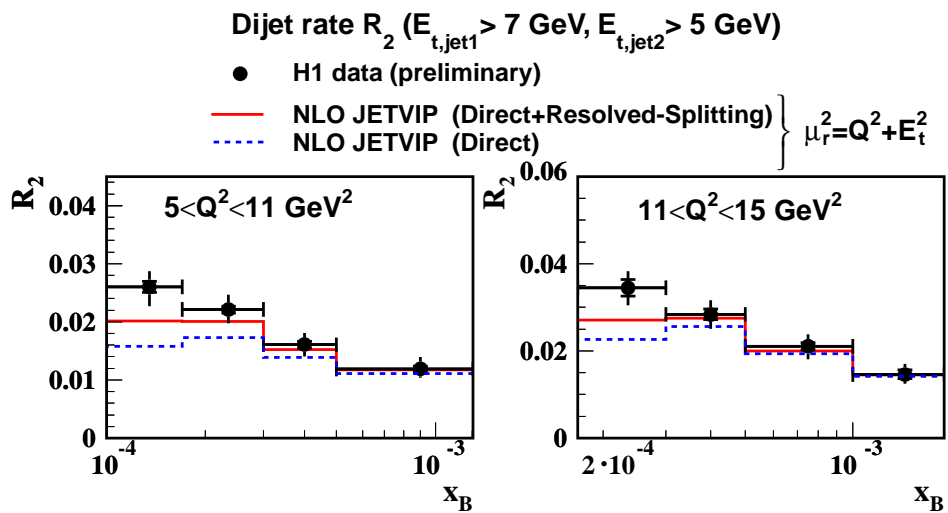,height=6.8cm}
\end{center}
\vspace{-0.5cm}
\caption{\label{fig:dijets_jetvip} Ratio of cross sections for dijet
and inclusive production as a function of $x_{\rm B}$. The data are
shown as the points with statistical errors (inner 
bars) and statistical and systematic uncertainties added in quadrature
(outer bars). The NLO calculation is shown with (solid line) and
without (dashed line) a resolved component.}
\end{figure}

In this section I will focus on a region of $Q^2$ which is not yet well
understood, i.e.\ the intermediate region between photoproduction
and DIS. This region involves photon virtualities roughly in the range
$1< Q^2 < 100$~GeV$^2$. 

In \cite{zz} a study was made of the scale dependence for the
NLO corrections to the dijet cross section employing the inclusive
$\kperp$ algorithm. The size of the scale variation is an indication
of the possible size of perturbative higher-order contributions. This
study is displayed in Fig.~\ref{fig:hadcoralgo}. Shown is the
$k$-factor, defined as the ratio of the NLO and the LO predictions,
for two different choices of the  renormalization scale ($\mu_r =
\avet, Q$). Towards low $Q^2$ the NLO corrections become large,
especially for the choice $\mu_r=Q$. Reasonably small $k$-factors
($k<1.4$) are only seen roughly at $Q^2 > 150$~GeV$^2$ where $Q^2$ and
$E_T^2$ are of similar size such that terms $\propto \ln(E_T^2 / Q^2)$
are small. The renormalization scale dependence is seen to be
correlated  with the NLO correction i.e.\  large at small $Q^2$. The
factorization scale dependence was found to be below 2\% over the
whole phase space. 

This study indicates that higher order (i.e.\ NNLO) corrections may
be important in the region of $Q^2<100$~GeV$^2$. One way of incorporating
these higher order contributions might be to include the resolved
virtual photon contribution. This is clearly a justified approach for
small $Q^2$ around 1~GeV$^2$, since here the photoproduction
regime starts. In photoproduction it is well established that the
resolved photon contribution is an important part of the jet cross
sections. The inclusion of a resolved photon into the calculation compared to
data from dijet production in the region $5<Q^2<15$~GeV$^2$ and at least 
two jets, such that again $E_T^2\gg Q^2$, is shown in
Fig.~\ref{fig:dijets_jetvip} (taken from \cite{wing}). It can be seen that  
the calculation lies below the data at low $x_{\rm B}$ and that the
description improves with the inclusion of a resolved photon. The
effect is larger at the lower range in $Q^2$, but is not enough to
completely describe the data. 

Another region of phase space where resolved virtual photon
contributions has been discussed to give important contributions is
that of the forward jet cross sections. The H1 and ZEUS collaborations
have measured forward jet cross sections at small $x$ for rather
similar kinematical conditions \cite{x5,x6}. The jet selection
criteria and kinematical cuts are summarized in Tab.~1.  In \cite{xkp}
we have performed a NLO calculation including the virtual 
resolved photon for the forward jet region  with the help of
{\tt JetViP}.

\begin{table}[bbb]
\caption{Forward jet selection criteria by H1 and ZEUS}
\begin{center}
\renewcommand{\arraystretch}{1.4}
\setlength\tabcolsep{5pt}
\begin{tabular}{cc}
\hline\hline
\makebox[4.cm][c]{H1 cuts} &  \makebox[4.cm][c]{ZEUS cuts}\\ \hline 
$E_{e}^{\prime} > 11$ GeV & $E_{e}^{\prime} > 10$ GeV \\
$y_e > 0.1$ & $y_e > 0.1$ \\
$E_{T,jet} > 3.5$ ($5$) GeV & $E_{T,jet} > 5$ GeV \\
$1.7 < \eta_{jet} < 2.8$ & $\eta_{jet} < 2.6$ \\ 
$0.5 < E_{T,jet}^{2} / Q^{2} < 2$ & $0.5 < E_{T,jet}^{2} / Q^{2} < 2$ \\ 
$x_{jet} > 0.035$ & $x_{jet} > 0.036$ \\ \hline \hline
\end{tabular}
\end{center}
\label{Tab1a}
\end{table}

\begin{figure*}[ttt]
  \unitlength1mm
  \begin{picture}(122,57)
    \put(27,-50){\epsfig{file=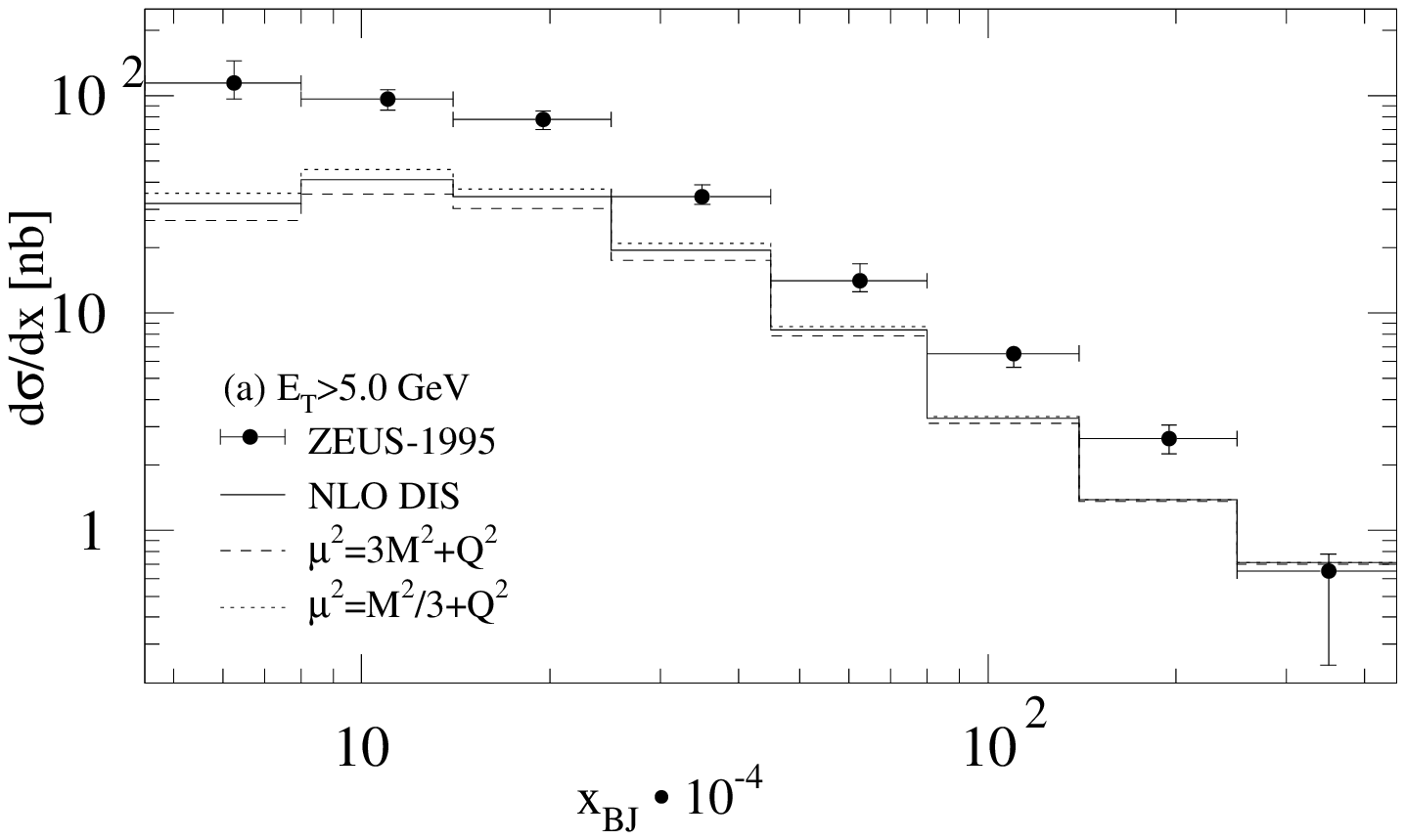,width=6.6cm,height=11cm}}
    \put(88,-50){\epsfig{file=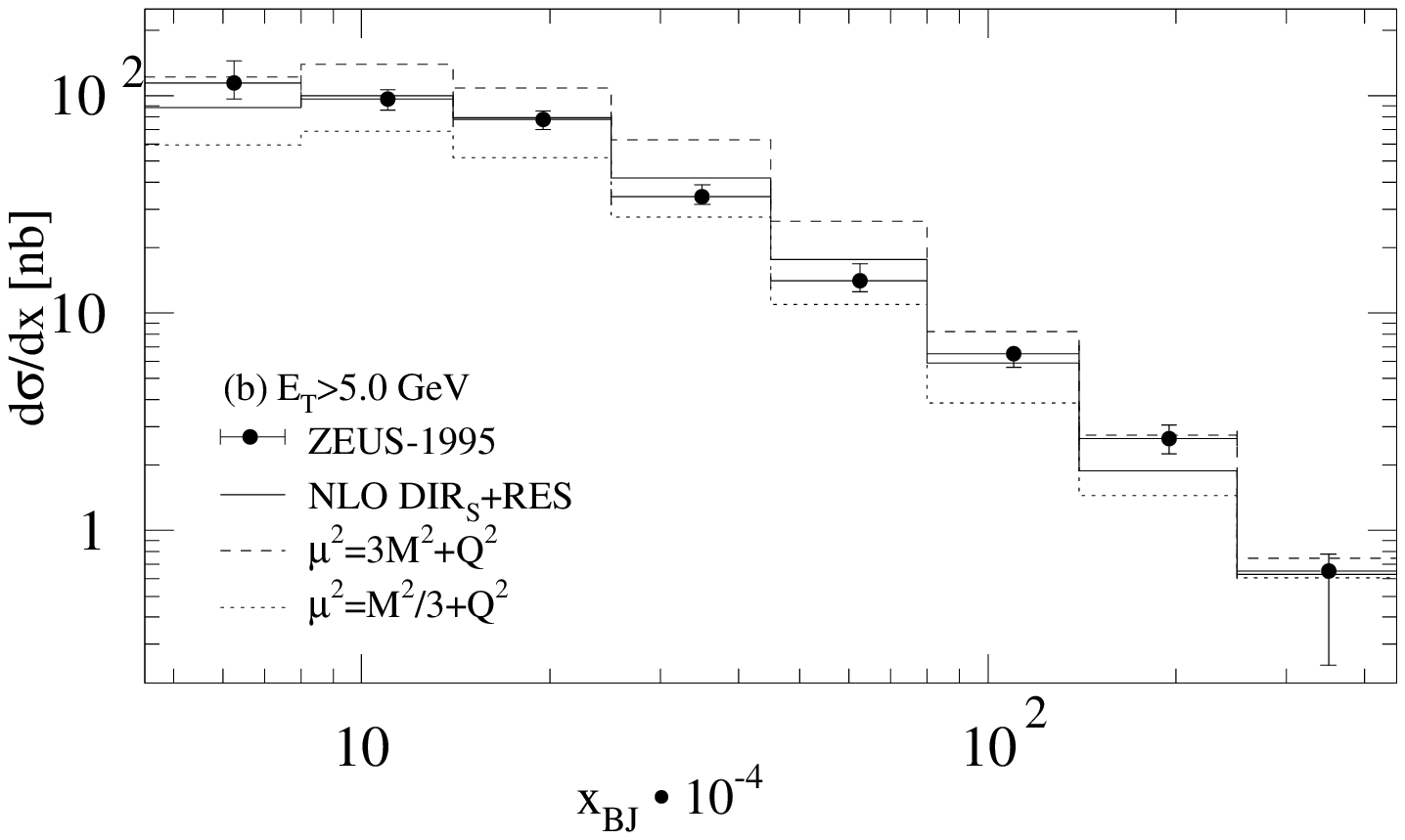,width=6.6cm,height=11cm}}
  \end{picture}
  \begin{picture}(122,51)
    \put(27,-50){\epsfig{file=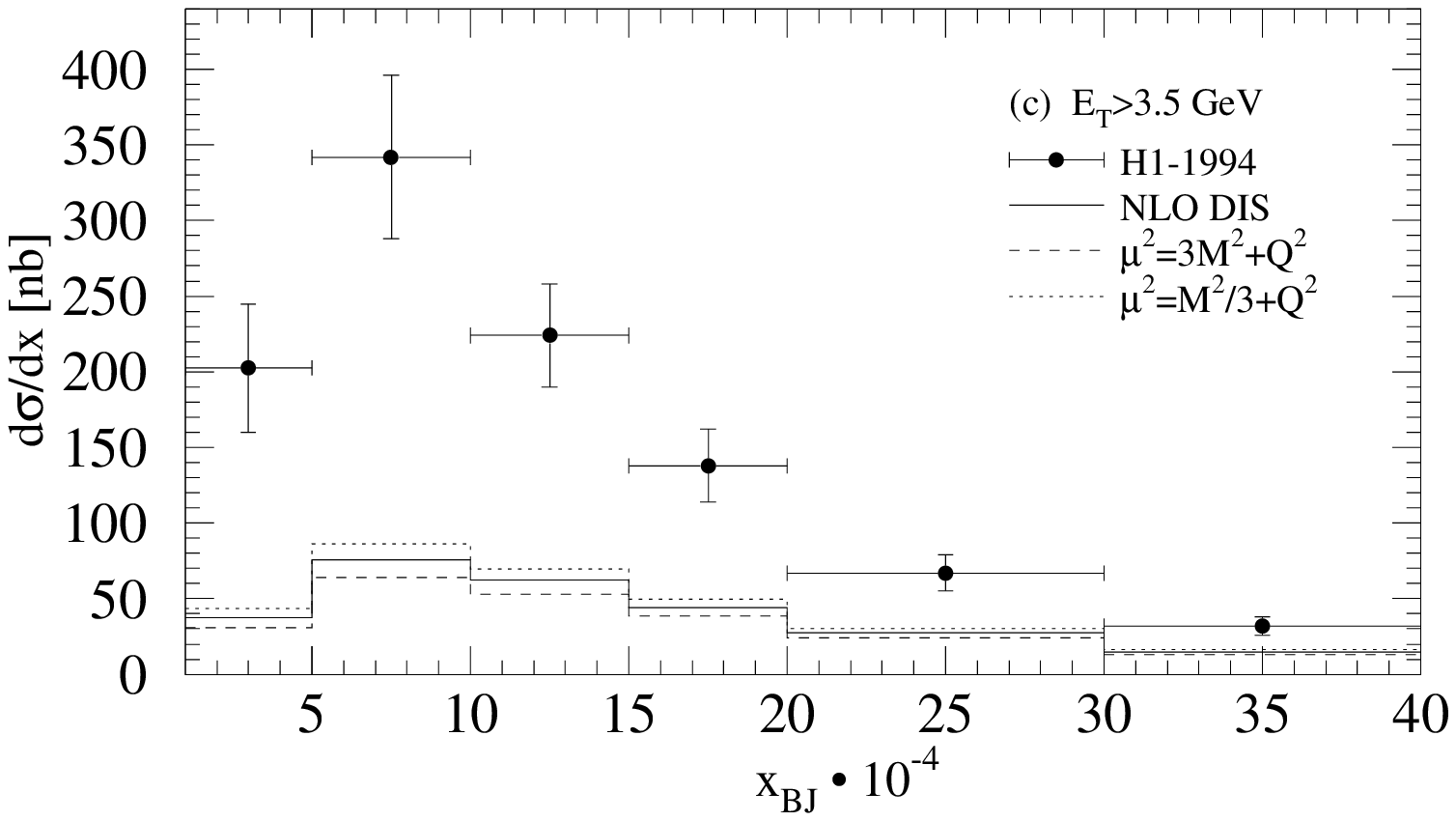,width=6.6cm,height=11cm}}
    \put(88,-50){\epsfig{file=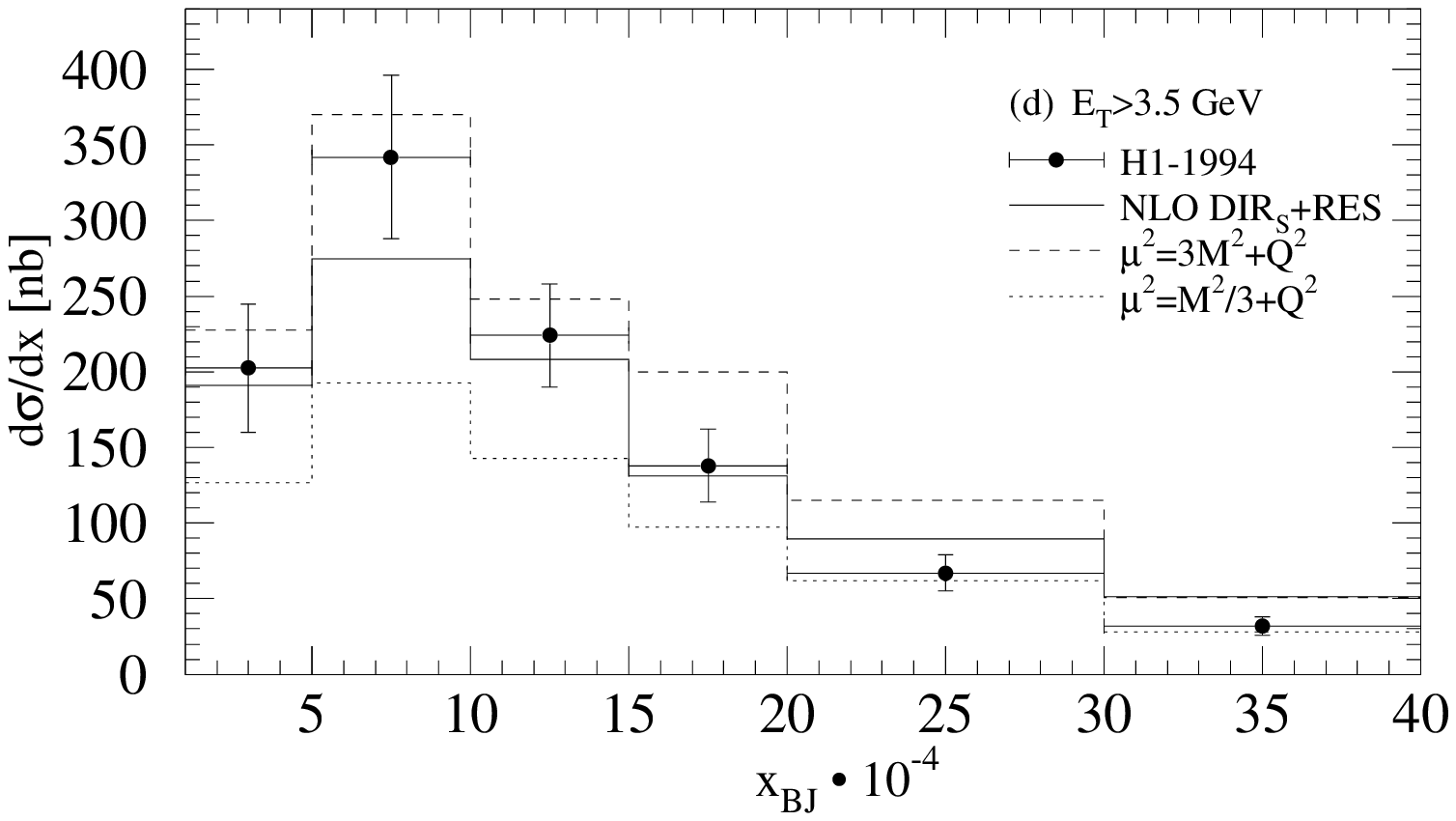,width=6.6cm,height=11cm}}
  \end{picture}
\caption{Dijet cross section in the forward region compared to HERA
data: (a) and (b) ZEUS; (c) and (d) H1.
(a) NLO DIS, $E_T>5$ GeV; (b) NLO DIR$_S$+RES, $E_T>5$ GeV;
(c) NLO DIS, $E_T>3.5$ GeV; (d) NLO DIR$_S$+RES, $E_T>3.5$ GeV.
\label{xkp1}}
\end{figure*}

The results for the ZEUS kinematical conditions are shown in
Fig.~\ref{xkp1}~a,b. In Fig.~\ref{xkp1}~a we plotted the full ${\cal
O}(\alpha_s^2)$ inclusive two-jet cross section (DIS) as a function
of $x$ for three different scales $\mu^2=\mu_R^2=3M^2+Q^2, M^2+Q^2$
and $M^2/3+Q^2$ with a fixed $M^2=50$~GeV$^2$ related to the mean
$E_T^2$ of the forward jet and compared them with the measured points
from ZEUS \cite{x5}. The choice $\mu_F^2 > Q^2$ is mandatory if we
want to include a resolved contribution. Similar to the results
obtained with MEPJET and DISENT, the NLO direct cross section is by a
factor 2 to 4 too small compared to the data. The variation inside the
assumed range of scales is small, so that also with a reasonable
change of scales we can not get agreement with the data. In
Fig.~\ref{xkp1}~b we show the corresponding forward jet cross sections
with the NLO resolved contribution included, labeled DIR$_S$+RES,
again for the  three different scales $\mu $ as in
Fig.~\ref{xkp1}~a. Now we find good agreement with  the ZEUS data. The
scale dependence is not so large that we must fear our results not to
be trustworthy.

In Fig.~\ref{xkp1}~c,d we show the results compared to the H1 data \cite{x6}
obtained with $E_T>3.5$~GeV in the HERA system. In the plot on the left 
the data are compared with the pure NLO direct prediction, which turns
out to be too small by a similar factor as observed in the comparison
with the ZEUS data. In Fig.~\ref{xkp1}~d the forward jet cross section is
plotted with the NLO resolved contribution included in the way
described above. We find good agreement with the H1 data inside
the scale variation window $M^2/3+Q^2<\mu^2 < 3M^2+Q^2$. We have also
compared the predictions with the data from the larger $E_T$ cut, namely
$E_T>5.0$~GeV, and found similar good agreement \cite{xkp}. The
enhancement of the NLO direct cross section through inclusion of
resolved processes in NLO is mainly due to the convolution of the
point-like term in the photon PDF with the NLO resolved matrix
elements, which gives an approximation to the NNLO direct cross
section without resolved contributions. One can therefore speculate
that the  forward jet cross section could be described within a fixed
NNLO calculation, using only direct photons.
In summary, the NLO theory with a resolved virtual photon
contribution as an approximation of the NNLO DIS cross section, which
is presently not available, gives a good description of the forward
jet data.

These results indicate that the resolved virtual photon approach might give
some higher order contributions. However, the concepts of
factorization and of the virtual photon structure function itself are
not well defined for too large photon virtualities (i.e.\ certainly
for $Q^2>10$~GeV$^2$) due to $k_\perp$ and mass effects. This is a
problem to be clearified in the future.

%**********************************************************************
%**********************************************************************
\section{Combining NLO QCD Calculations and Parton Showers}  

There are two approaches to the problem of simulating high-energy
physics processes. First, one can employ fixed order QCD calculations which  
deliver the partonic final state of a single event and which currently 
are available up to next-to-leading order (NLO) for the case of inclusive
jet production at HERA. This approach has been discussed so
far. Second, Monte Carlo models are widely used which implement
phenomenological descriptions of the parton cascades in the initial
and final states, like for example parton shower (PS) algorithms in
which a resummation of leading $\log Q^2$ terms to all orders and a
natural matching to hadronisation models lead to a good description of
small-angle phenomena (see \cite{mcworkshop} for a summary of programs
in $eP$-scattering).

The technical realisation of parton shower models has several 
shortcomings: A beforehand-calculation of the total cross-section is
required to get a correct normaliztion of the cross sections, which
might be difficult for any other physics scenario than the one-jet
case. The cross-section predictions in the 'soft' parton
shower region as well as in the 'hard' region are only in leading order (LO). 
Going to NLO, which is necessary in order to increase the predictive power
of the event generators, is however made difficult by negative cross-section 
contributions which lead to numerical stability problems when being combined
with the probabilistic Sudakov approach to parton showers. In order to
improve on the predictive power of the simulations, we have developed
\cite{cm1} and implemented \cite{cm2} a method with which negative
weights are avoided during the NLO calculation such that one easily
arrives at the correct NLO weight in the PS region. For other
approaches for combining PS and fixed order calculations, see \cite{other}.

The Born term $\sigma^B$ and the virtual corrections $\sigma^V$ to
the NLO cross-section are easily evaluated analytically; the real
corrections $\sigma^R$ can be written as an integral from $0$ to
$R_{tech}$ over a function $F(z)$ which is basically the matrix
element without the propagator (beyond $R_{tech}$ the hard region
begins). One can then split the real corrections integral by some
arbitrary parameter $\delta$. Choosing this parameter $\delta$ such
that the Born term, the virtual corrections and the soft/collinear
part of the real corrections (the two-body parts) exactly cancel  
we are left only with an integral over the finite part of the real corrections
(three-body contributions) which gives the complete NLO cross-section
in the parton shower region. The corresponding value of $\delta$,
which depends on $x$ and $Q^2$ and the renormalisation and factorisation
scales, is denoted by $\tilde{\delta}$:
\begin{equation}
 \sigma^{NLO}_{PS} = \sigma^B + \sigma^V + \sigma^R \simeq
 \int_{\tilde{\delta}}^{R_{tech}}\frac{dz}{z}F(z) > 0.
\label{eq1}
\end{equation}

For the inclusive one-jet cross-section calculations we implemented \cite{cm2}
the analytically derived function $\tilde{\delta}$ in the NLO QCD program 
{\tt DISENT} \cite{1} and combined the NLO cross-sections thus obtained
for the PS region with the final state parton shower from 
{\tt PYTHIA} \cite{pythia}. Events where $2p_ip_j < \tilde{\delta}Q^2$ 
for any pair of partons $i,j$ are rejected (this is the soft/collinear 
emission region). The hard region is described by $\mathcal{O}(\alpha_s)$ 
matrix elements. PS emissions into the hard region are vetoed in order 
to avoid double-counting. The parameter $R_{tech}$ is typically of the
order of 1. We have checked that our results are reasonably stable
against a variation of $R_{tech}$ by a factor of 2. 

\begin{figure}[t]
\unitlength1mm
\begin{picture}(122,125)
  \put(17,0){\epsfig{file=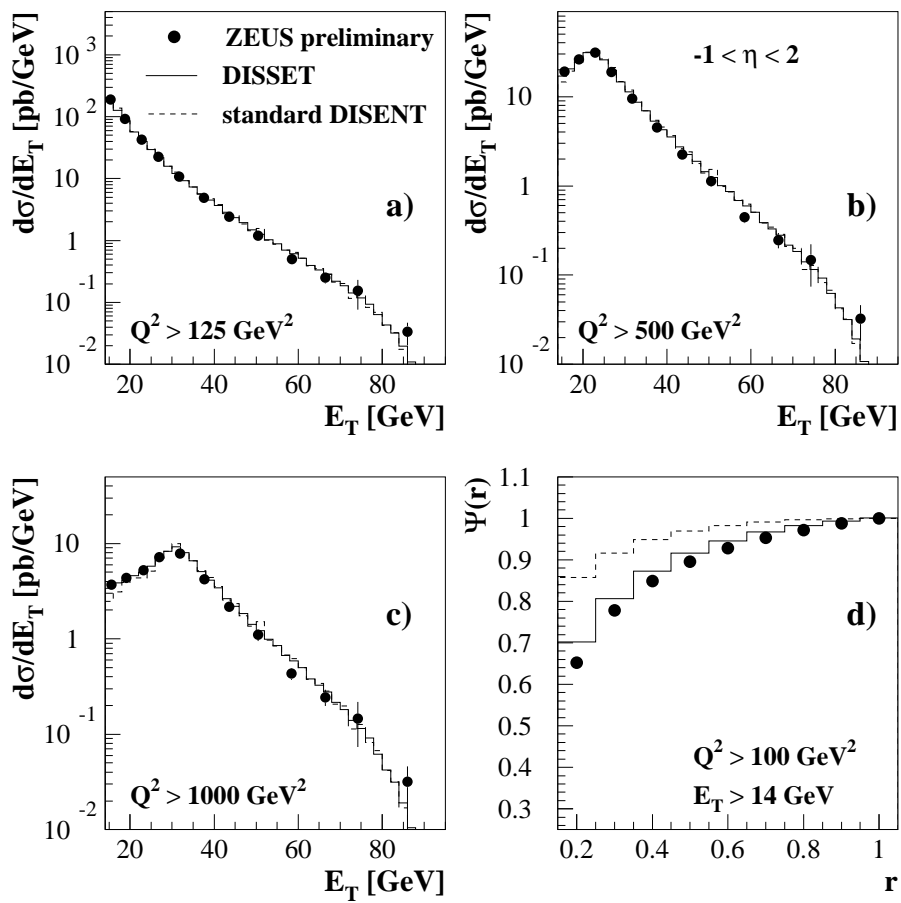,width=12cm}}
\end{picture}
\caption{$E_T$ distributions in different bins of $Q^2$ (a-c) and
integrated jet shape (d). Shown are ZEUS data, standard {\tt DISENT}
NLO results and results of our calculation called {\tt DISSET}.\label{fig1}} 
\end{figure}

The ZEUS collaboration at HERA has published jet spectra \cite{zeus1} 
and jet shape data \cite{zeus2} for the case of inclusive jet 
production in the laboratory frame. For the spectra, events with $Q^2
>$125 $\mathrm{GeV^2}$, $y<0.95$ and an energy of the scattered
electron of more than 10 GeV were selected. The jets which were
reconstructed using a $k_\perp$ algorithm were required to have
transverse energies $E_T>$14 GeV and a pseudorapidity
-1$<\eta<$2. For the jet shape analysis, an iterative cone jet
algorithm was used, and $Q^2>$100$\mathrm{GeV^2}$ was required.

In Fig.~\ref{fig1} we show comparisons of ZEUS inclusive jet $E_T$
spectra in three bins of $Q^2$ (a-c) and of the integrated jet shape
for the full $Q^2$ and $E_T$ ranges (d) with a standard {\tt DISENT}
NLO calculation and with our results (called {\tt DISSET}). 
Good agreement between the data, our result and {\tt DISENT} 
can be observed for the inclusive spectra. For the jet shapes the {\tt
DISENT} result, which is only LO, fails. Our result however is much
closer to the data since we describe the soft region by the parton
shower. The remaining difference between the data and our predictions
can be accounted for by hadronisation effects as  we checked using the
{\tt LEPTO} event generator\cite{lepto}. These results show  
that we correctly combine the NLO cross-section normalisation
with the parton shower which describes the details of 
the hadronic final such as the jet shapes.

Currently, we are investigating the technically much more complicated
case of dijet production in the Breit frame. In principle the method 
described above works also for this physically more relevant case.
However it turns out that the analytical solution of equation
(\ref{eq1}) is not straight forward. The relevant equation to be solved
for $\tilde{\delta}$ becomes
\begin{equation}
0=\sum_{i=0}^2\ln(\tilde{\delta})^i\cdot A_i +
A_3\cdot\ln(1+f(s,t,u,\xi)/\tilde {\delta}) +
A_4\cdot\ln(1+g(s,t,u,\xi)/\tilde{\delta}) 
\end{equation}
with analytical expressions for coefficients $A_i$. We decided for a
numerical evaluation of this equation using Newton's method which
after 4 to 5 iterations gives stable results, such that this way of
estimating the cut-off is applicable in a MC enviroment, where many
events have to be generated.

%**********************************************************************
%**********************************************************************
\section{Summary}

We have described the status of next-to-leading order calculations for
jet production in $eP$-scattering at HERA. We discussed the
photoproduction, DIS and intermediate regimes and found that
especially the range of intermediate photon virtualities
$1<Q^2<100$~GeV$^2$ to be not
well described by present NLO calculations. We finally described
progress made for the problem of including higher order corrections
into MC models, that include PS and hadronization.

%**********************************************************************
%**********************************************************************

\end{document}